\documentclass{article}
\usepackage{ijcai17}

\usepackage{times}
\usepackage{array}
\usepackage{amssymb}
\usepackage{amsmath}
\usepackage{algorithm}
\usepackage{latexsym}
\usepackage{proof}
\usepackage[noend]{algpseudocode}
\usepackage{fancyvrb}
\usepackage{verbatim}
\usepackage{alltt}
\usepackage{galois}
\usepackage{multirow}
\usepackage{mathtools} 
\usepackage{url}

\usepackage{listings}
\usepackage{bm}

\usepackage{amsthm}
\usepackage{thmtools}

\title{Synthesizing Imperative Programs for Introductory Programming Assignments}
\author{Sunbeom So \\ 
Korea University \\
sunbeom\_so@korea.ac.kr \\ 
\And 
Hakjoo Oh \\
Korea University \\
hakjoo\_oh@korea.ac.kr}

\begin{document}
\newcommand{\union}{\cup}
\newcommand{\infinity}{\infty}

\newcommand{\longversion}{false}
\newcommand{\anonymous}{false}
 \newcommand{\hide}[1]{}

\newcommand{\cP}{\mathcal{P}}
\newcommand{\cQ}{\mathcal{Q}}
\newcommand{\cH}{\mathcal{H}}
\newcommand{\RA}{\Rightarrow}

\newcommand{\db}[1]{\llbracket #1 \rrbracket}
\newcommand{\dbpre}[1]{\llbracket #1 \rrbracket^{\pre}}
\newcommand{\analysis}{\mathsf{analyze}}
\newcommand{\prune}{{\sf prune}}

\newcommand{\mbit}[1]{\mbox{\it #1}}
\newcommand{\mbrm}[1]{\mbox{\rm #1}}
\newcommand{\mbtt}[1]{\mbox{\tt #1}}
\newcommand{\mbsf}[1]{\mbox{\sf #1}}
\newcommand{\sem}[1]{[\![#1]\!]}
\newcommand{\sema}[1]{\mathcal{A}[\![#1]\!]}
\newcommand{\semb}[1]{\mathcal{B}[\![#1]\!]}
\newcommand{\semc}[1]{\mathcal{C}[\![#1]\!]}
\newcommand{\abssemp}{\mathcal{P}[\![p]\!]}
\newcommand{\trunc}[1]{||#1||}

\newcommand{\edge}{\hookrightarrow}
\newcommand{\intraedge}{\edge_{\mbit{f}}}
\newcommand{\calledge}{\edge_{\mbit{c}}}
\newcommand{\returnedge}{\edge_{\mbit{r}}}
\newcommand{\cedge}{\edge_{\mbit{cfg}}}
\newcommand{\defeq}{\triangleq}
\newcommand{\aedge}{\hat{\hookrightarrow}}
\newcommand{\aintraedge}{\aedge_{\mbit{f}}}
\newcommand{\acalledge}{\aedge_{\mbit{c}}}
\newcommand{\areturnedge}{\aedge_{\mbit{r}}}
\newcommand{\finto}{\stackrel{\textsf{fin}}{\to}}
\newcommand{\join}{\sqcup}
\newcommand{\lub}{\bigsqcup}
\newcommand{\ple}{\sqsubseteq}
\newcommand{\meet}{\sqcap}
\newcommand{\bigjoin}{\bigsqcup}
\newcommand{\bigmeet}{\bigsqcap}

\newcommand{\mca}{\mathcal{A}}
\newcommand{\mcb}{\mathcal{B}}
\newcommand{\mcc}{\mathcal{C}}
\newcommand{\mcf}{\mathcal{F}}
\newcommand{\mcg}{\mathcal{G}}
\newcommand{\mcp}{\mathcal{P}}
\newcommand{\mcl}{\mathcal{L}}
\newcommand{\mcu}{\mathcal{U}}
\newcommand{\mci}{\mathcal{I}}
\newcommand{\mcv}{\mathcal{V}}

\newcommand{\run}{\mcf}

\newcommand{\Control}{\mbit{Control}}
\newcommand{\Context}{\mbit{Context}}
\newcommand{\FunSymb}{\mbit{FunSymb}}
\newcommand{\Table}{\mbit{Table}}
\newcommand{\CallSite}{\mbit{CallSite}}
\newcommand{\ReturnSite}{\mbit{ReturnSite}}
\newcommand{\Itv}{\hat{\mathbb{Z}}}
\newcommand{\aItv}{\tilde{\mathbb{Z}}}
\newcommand{\push}{\mbsf{push}}
\newcommand{\pop}{\mbsf{pop}}
\newcommand{\funSymb}{\mbsf{funSymb}}
\newcommand{\lfp}{\mbsf{lfp}}
\newcommand{\fix}{\mbsf{fix}}
\newcommand{\afix}{\myhat{\mbsf{fix}}}
\newcommand{\suc}{\mbsf{succ}}
\newcommand{\assume}{\mbsf{assume}}
\newcommand{\callee}{\mbsf{callee}}
\newcommand{\lang}{\mcl}
\newcommand{\eval}{\mbit{eval}}
\newcommand{\nondet}{\mbtt{nondet}}
\newcommand{\ite}{\mbrm{ite}}
\newcommand{\Def}{\mbsf{D}}
\newcommand{\Use}{\mbsf{U}}
\newcommand{\widenop}{\bigtriangledown}
\newcommand{\Meta}{\mbit{Meta}}
\newcommand{\Precision}{\mbit{Precision}}
\newcommand{\Trend}{\mbit{Trend}}
\newcommand{\Sign}{\mbit{Sign}}
\newcommand{\Lower}{\mbit{L}}
\newcommand{\Upper}{\mbit{U}}
\newcommand{\Inc}{\mbit{Inc}}
\newcommand{\Dec}{\mbit{Dec}}
\newcommand{\swap}{\mbsf{swap}}
\newcommand{\inc}{\mbit{inc}}
\newcommand{\dec}{\mbit{dec}}
\newcommand{\myneg}{\mbit{neg}}
\newcommand{\myassume}{\mbtt{assume}}
\newcommand{\reset}{\mbtt{reset}}
\newcommand{\dom}{\mbrm{dom}}
\newcommand{\trend}{\mbit{trend}}
\newcommand{\widen}{\mbit{widen}}
\newcommand{\branch}{\,[\!]\,}

\newcommand{\intcomps}{R_{int}}
\newcommand{\lvcomps}{R_{lv}}
\newcommand{\varcomps}{R_{ivar}}
\newcommand{\arrcomps}{R_{avar}}

\newcommand{\atrans}{\hat{F}}
\newcommand{\watrans}{\atrans_w}
\newcommand{\pstrans}{F_{PS}}
\newcommand{\ptrans}{F_P}
\newcommand{\strans}{F_S}
\newcommand{\ttrans}{F_T}
\newcommand{\mtrans}{F_{m}}

\newcommand{\aeval}{\eval}
\newcommand{\peval}{\eval_P}
\newcommand{\seval}{\eval_S}
\newcommand{\meval}{\eval_{m}}

\newcommand{\mba}{\mathbb{A}}
\newcommand{\mbb}{\mathbb{B}}
\newcommand{\mbc}{\mathbb{C}}
\newcommand{\mbd}{\mathbb{D}}
\newcommand{\mbe}{\mathbb{E}}
\newcommand{\mbf}{\mathbb{F}}
\newcommand{\mbg}{\mathbb{G}}
\newcommand{\mbh}{\mathbb{H}}
\newcommand{\mbi}{\mathbb{I}}
\newcommand{\mbj}{\mathbb{J}}
\newcommand{\mbk}{\mathbb{K}}
\newcommand{\mbl}{\mathbb{L}}
\newcommand{\mbm}{\mathbb{M}}
\newcommand{\mbn}{\mathbb{N}}
\newcommand{\mbo}{\mathbb{O}}
\newcommand{\mbp}{\mathbb{P}}
\newcommand{\mbq}{\mathbb{Q}}
\newcommand{\mbr}{\mathbb{R}}
\newcommand{\mbs}{\mathbb{S}}
\newcommand{\mbt}{\mathbb{T}}
\newcommand{\mbu}{\mathbb{U}}
\newcommand{\mbv}{\mathbb{V}}
\newcommand{\mbw}{\mathbb{W}}
\newcommand{\mbx}{\mathbb{X}}
\newcommand{\mby}{\mathbb{Y}}
\newcommand{\mbz}{\mathbb{Z}}

\newcommand{\cfgto}{\hookrightarrow}
\newcommand{\calls}{\rightarrowtail}
\newcommand{\returns}{\dashrightarrow}

\newcommand{\fid}{\mathsf{fid}}
\newcommand{\callof}{\mathsf{callof}}
\newcommand{\retof}{\mathsf{retof}}
\newcommand{\entryof}{\mathsf{entryof}}
\newcommand{\exitof}{\mathsf{exitof}}
\newcommand{\mysucc}{\mathsf{succ}}
\newcommand{\mypred}{\mathsf{pred}}
\newcommand{\callstring}{\mathsf{callstring}}
\newcommand{\hd}{\mathsf{hd}}
\newcommand{\tl}{\mathsf{tl}}

\newcommand{\floor}[1]{\lfloor #1 \rfloor}
\newcommand{\ceil}[1]{\lceil #1 \rceil}

\newcommand{\zap}[1]{}
\newcommand{\ctxgraph}{\mcg}
\newcommand{\fctxgraph}{\ctxgraph_{\fcfa}}
\newcommand{\nodes}{N}
\newcommand{\node}{n}
\newcommand{\edges}{E}
\newcommand{\pair}[2]{\langle #1, #2 \rangle}
\newcommand{\myset}[1]{\{ #1 \}}
\newcommand{\Index}{{\Delta}}
\newcommand{\Var}{\mbx}
\newcommand{\iVar}{\Var_{\it i}}
\newcommand{\aVar}{\Var_{\it a}}
\newcommand{\Sym}{\mathsf{Sym}}
\newcommand{\arr}[1]{\langle #1 \rangle}
\newcommand{\sfclos}[1]{#1_{\rightarrow}}
\newcommand{\fcfa}{{\infty\mathsf{cfa}}}

\newcommand{\Reach}{\mathcal{R}}
\newcommand{\VFGNode}{\mathcal{N}}
\newcommand{\Preserve}{\mathcal{P}}
\newcommand{\aQuery}{{\Query}^{\pre}}
\newcommand{\solp}{\mathcal{X}}
\newcommand{\solpre}{\mathsf{PA}}
\newcommand{\solmain}{\mathsf{MA}}
\newcommand{\abst}{}
\newcommand{\lnontop}{\bigstar}
\newcommand{\ltop}{\top_v}
\newcommand{\lbot}{\bot_v}

\newcommand{\lv}{\it lv}
\newcommand{\e}{e}
\newcommand{\assignment}[2]{#1:=#2}
\newcommand{\skipcmd}{{\it skip}}
\newcommand{\store}[2]{#1:=#2}
\newcommand{\assert}[2]{\{\!\!\{#1<#2\}\!\!\}}
\newcommand{\call}[3]{\mbox{\it call}(#1_{#2},#3)}
\newcommand{\returncmd}{{\it return}}
\newcommand{\aState}{\mbs}
\newcommand{\aV}{\abst{\mbv}}
\newcommand{\atransitpart}{\hookrightarrow}
\newcommand{\aG}{\abst{G}}
\newcommand{\aX}{\abst{X}}
\newcommand{\af}{\abst{f}}
\newcommand{\as}{\abst{s}}
\newcommand{\cmd}{{\it cmd}}
\newcommand{\subst}[1]{[ #1 ]}
\newcommand{\evalexp}[1]{\hat{\mathcal{E}}(#1)}
\newcommand{\evalexppre}[1]{\hat{\mathcal{E}^{\pre}}(#1)}
\newcommand{\num}       [1][n]  {#1}
\newcommand{\defusea}{\rightsquigarrow}
\newcommand{\defusex}{\sim}
\newcommand{\defuse}[1][l]{\stackrel{#1}{\defusea}}
\newcommand{\Paths}{{\sf Paths}}
\newcommand{\VPaths}{\mbox{\sf VPaths}}
\newcommand{\SPaths}{\mbox{\sf SPaths}}
\newcommand{\DPaths}{\mbox{\sf DPaths}}
\newcommand{\PrevPaths}{\mbox{\sf slice}}
\newcommand{\nil}{\varnothing}
\newcommand{\power}[1]{\wp(#1)}

\newcommand{\vflow}{\succ}

\newcommand{\mydot}{.\hspace{1pt}}
\newcommand{\st}{\mydot}
\newcommand{\tup}[1]{\langle#1\rangle}

\newcommand{\slice}{\backslash}
\newcommand{\mystar}{*}
\newcommand{\myland}{\;\land\;}
\newcommand{\mylor}{\;\vee\;}

\newcommand{\atransit}{\cfgto}
\newcommand{\ctx}{\gamma}

\newcommand{\callgraph}{\mathsf{cg}}

\newtheorem{conjecture}{Conjecture}
\newtheorem{proposition}{Proposition}
\newtheorem{corollary}{Corollary}
\newtheorem{assumption}{Assumption}
\newtheorem{example}{Example}
\newtheorem{theorem}{Theorem}
\newtheorem{lemma}{Lemma}
\newtheorem{definition}{Definition}

\newcommand{\hold}{\mathsf{hold}}
\newcommand{\holdpre}{\mathsf{hold}^{\pre}}
\newcommand{\bool}{\mathsf{bool}}
\newcommand{\avpre}{\av^{\pre}}
\newcommand{\av}{\hat{v}}

\newcommand{\todoc}[2]{{\textcolor{#1} {\textbf{[[#2]]}}}}

\newcommand{\todored}[1]{\todoc{red}{#1}}
\newcommand{\todoblue}[1]{\todoc{blue}{#1}}
\newcommand{\todogreen}[1]{\todoc{green}{#1}}

\newcommand{\TODO}[1]{\todored{#1}}
\newcommand{\hy}[1]{\todoblue{HY: #1}}
\newcommand{\hj}[1]{\todored{HO: #1}}

\newcommand{\argmax}{\operatornamewithlimits{argmax}}
\newcommand{\cardinality}[1]{\left\vert{#1}\right\vert}
\let\oldvec\vec
\renewcommand{\vec}[1]{\mathbf{#1}}
\newcommand{\component}{\mbj}
\newcommand{\idx}{j}
\newcommand{\pgm}{\mbp}
\newcommand{\Query}{\mbq}

\newcommand{\oracle}{{\mathcal{S}}}
\newcommand{\classifier}{{\mathcal{C}}}
\newcommand{\strategy}{{\mathcal{S}}}

\newcommand{\trainset}{\mathcal{T}}
\newcommand{\trainsetp}{\mathcal{T}_1}
\newcommand{\trainsetn}{\mathcal{T}_0}

\newcommand{\score}{{\it score}}
\newcommand{\wv}{\vec{w}}

\newcommand{\ActiveCoarsen}{\textsc{ActiveCoarsen}}
\newcommand{\maa}{\textsc{MinAbs}}
\newcommand{\ma}{\vec{a}_{\mathit{min}}}

\newcommand{\model}{\mathcal{M}}
\newcommand{\data}{\theta}
\newcommand{\Pgm}{\mbp}
\newcommand{\Procs}{\mathit{Procs}}
\newcommand{\myPr}{\mathit{Pr}}
\newcommand{\Sparrow}{\textsc{Sparrow}}

\newcommand{\bs}{{\bf s}}
\newcommand{\bk}{{\bf k}}
\newcommand{\bK}{{\bf K}}

\newcommand{\myand}{\land}
\newcommand{\myor}{\vee}
\newcommand{\mynot}{\lnot}

\newcommand{\reducer}{{\sf reduce}}
\newcommand{\cond}{{{\sf cond}}}
\newcommand{\define}{=}
\newcommand{\assertChecker}{{\sf check}}
\newcommand{\pdtrans}{{\sf impair}}

\newcommand{\deppath}{P}
\newcommand{\acmd}{\hat{\it c}}
\newcommand{\aexp}{\hat{\it e}}
\newcommand{\abinop}{\oplus}
\newcommand{\bbinop}{\prec}
\newcommand{\compare}{\prec}

\newcommand{\alv}{\hat{lv}}

\newcommand{\StaticAnalyzer}{F}
\newcommand{\Abs}{\mathcal{A}}

\newcommand{\fgen}{{\sf extract}}
\newcommand{\fpred}{{\sf match}}

\newcommand{\reducecond}{\phi}

\newcommand{\assumecmd}{{\it assume}}
\newcommand{\alloccmd}{{\it alloc}}

\newcommand{\featlang}{\mbf}
\newcommand{\feat}{\pi}
\newcommand{\Feat}{\Pi}

\newcommand{\query}{\mbq}
\newcommand{\threshold}{{\it t}}
\newcommand{\dep}{{\sf req}}
\newcommand{\predict}{{\sf predict}}

\newcommand{\pred}{{\sf pred}}

\newcommand{\solution}{\mathsf{solution}}
\newcommand{\hopeless}{\mathsf{hopeless}}
\newcommand{\analyzer}{\mathsf{possible}}
\newcommand{\final}{\mathsf{final}}
\newcommand{\false}{{\it false}}
\newcommand{\true}{{\it true}}
\newcommand{\normalize}{\mathsf{normalize}}
\newcommand{\overapprox}[1]{\widehat{#1}}
\newcommand{\underapprox}[1]{\widehat{#1}}
\newcommand{\nextstates}{{\mathsf{next}}}
\newcommand{\optimize}{{\mathsf{optimize}}}
\newcommand{\maxs}{{\mathsf{max}}}
\newcommand{\mins}{{\mathsf{min}}}
\newcommand{\holes}{{\mathsf{holes}}}
\newcommand{\subcand}{{\mathsf{split}}}
\newcommand{\trans}{{\mathsf{unroll}}}
\newcommand{\init}{{\mathsf{Init}}}
\newcommand{\states}{S}
\newcommand{\initialstate}{s_0}
\newcommand{\finalstates}{F}
\newcommand{\cost}{\mathcal{C}}

\newcommand{\even}{\mathsf{even}}
\newcommand{\odd}{\mathsf{odd}}
\newcommand{\sq}{\textsf{square}}

\newcommand{\ahole}{\lozenge}
\newcommand{\bhole}{\triangle}
\newcommand{\chole}{\square}

\newcommand{\dead}{{\sf fail}}
\newcommand{\range}{{\sf Ran}}
\newcommand{\SIMPL}{{\textsc{Simpl}}}
\newcommand{\SKETCH}{{\textsc{Sketch}}}
\newcommand{\PSKETCH}{{\textsc{PSketch}}}
\newcommand{\inputarr}{{\sf i_{arr}}}
\newcommand{\inputint}{{\sf i_{int}}}

\newcommand{\todo}{{\color{blue}{Todo}}}

\newcommand{\ifcmd}{{\it if}}
\newcommand{\whilecmd}{{\it while}}

\newcommand{\toa}{\to_a}
\newcommand{\tob}{\to_b}
\newcommand{\toc}{\to_c}

\newcommand{\integers}{\Gamma}
\newcommand{\vars}{\Sigma}
\newcommand{\ivars}{\vars_i}
\newcommand{\avars}{\vars_a}

\newcommand{\nvocab}{\Sigma}
\newcommand{\xvocab}{\Gamma}

\newcommand{\inv}{i}
\newcommand{\outv}{o}

\newcommand{\examples}{\mathcal{E}}
\newcommand{\resources}{\mathcal{R}}

\newcommand{\hole}{?}

\newcommand{\multiple}{\overrightarrow}

\newcommand{\sym}{\beta}
\newcommand{\se}{{\it se}}
\newcommand{\SE}{\mathsf{SE}}

\newcommand{\myhat}{\widehat}
\newcommand{\aMem}{\myhat{\mbm}}
\newcommand{\aValue}{\myhat{\mbv}}
\newcommand{\aBool}{\myhat{\mbb}}
\newcommand{\amem}{\myhat{m}}
\newcommand{\avalue}{\myhat{v}}
\newcommand{\aabinop}{\;\myhat{\abinop}\;}
\newcommand{\abbinop}{\;\myhat{\bbinop}\;}
\newcommand{\abssema}[1]{\myhat{\mathcal{A}}[\![#1]\!]}
\newcommand{\abssemb}[1]{\myhat{\mathcal{B}}[\![#1]\!]}
\newcommand{\abssemc}[1]{\myhat{\mathcal{C}}[\![#1]\!]}
\newcommand{\aland}{\;\myhat{\land}\;}
\newcommand{\avee}{\;\myhat{\vee}\;}
\newcommand{\aneg}{\myhat{\neg}}
\newcommand{\atrue}{\myhat{\true}}
\newcommand{\afalse}{\myhat{\false}}
\newcommand{\acond}{\myhat{\cond}}
\newcommand{\aF}{\myhat{F}}
\newcommand{\myskip}[1]{}


\maketitle
\begin{abstract}
  We present a novel algorithm that synthesizes imperative programs
  for introductory programming courses. Given a set of input-output 
  examples and a partial program, 
  our algorithm generates a complete program that is
  consistent with every example. Our key idea is to combine
  enumerative program synthesis and static analysis, 
  which aggressively prunes out a large search space while
  guaranteeing to find, if any, a correct solution.
  We have implemented our algorithm in a tool,
  called $\SIMPL$, and evaluated it on 30 problems used in
  introductory programming courses. The results show that $\SIMPL$ is
  able to solve the benchmark problems in 6.6 seconds on average.
\end{abstract}

\section{Introduction}

Our long-term goal is to build an intelligent tutoring system that
helps students to improve their programming skills.  Our experience in
introductory programming courses is that students, who learn
programming for the first time, often struggle with solving
programming problems for themselves.  Manually providing guidance
simply does not scale for the increasingly large number of students.
To make matters worse, we found that even instructors sometimes make
mistake and shy students are reluctant to ask questions. Motivated by
this experience, we aim to build an automatic system that helps
students to improve their skills without human teachers.

In this paper, we present a key component of the system, which
automatically generates complete programs from students' incomplete
programs.  
The inputs of the algorithm are a partial program 
with constraints on variables and constants, 
and input-output examples that specify the program's behavior.
The output is a complete program 
whose behavior matches all of the given input-output
examples.

The key novelty of our algorithm is to combine enumerative program
synthesis and program analysis techniques. It basically enumerates every possible
candidate program in increasing size until it finds a solution. This
algorithm, however, is too slow to be interactively used with students
due to the huge search space of programs. Our key idea to accelerate
the speed is to perform static analysis alongside the enumerative search, in order
to ``statically'' identify and prune out interim programs that
eventually fail to be a solution. We formalize our pruning technique
and its safety property. 

The experimental results show that our algorithm is remarkably
effective to synthesize introductory imperative programs. 
We have implemented the algorithm in a tool, $\SIMPL$, and evaluated
its performance on 30 programming tasks used in introductory
courses.
With our pruning technique, $\SIMPL$ is fast
enough to solve each problem in 6.6 seconds on average.  However,
without the pruning, the baseline algorithm, which already adopts
well-known optimization techniques, takes 165.5 seconds (25x slowdown) 
on average.

We summarize our contributions below:
\begin{itemize}
\item We present a new algorithm for synthesizing imperative programs
  from examples. To our knowledge, our work is the first to combine
  enumerative program synthesis and static analysis technologies.

\item We prove the effectiveness of our algorithm on 30 real
  programming problems used in introductory courses. The results show
  that our algorithm quickly solves the problems, including ones that most beginner-level students have hard times to solve.

\item We provide a tool, $\SIMPL$, which is publicly
  available and open-sourced.\footnote{Hidden for double-blind reviewing.}


\end{itemize}



\begin{figure}[t]
\centering
\begin{tabular}{c@{\quad}l@{\quad\quad}r}
\begin{minipage}{1.5in}


\begin{lstlisting}
$\textbf{reverse}$(n){
  r := 0; 
  while ($\fbox{\parbox[c][0.5em][c]{0.25\textwidth}{n > 0}}$){
    $\fbox{\parbox[c][4.0em][c]{0.60\textwidth}{
    x := n \% 10; \\
    r := r * 10; \\
    r := r + x; \\
    n := n / 10;  }}$
  }; 
  return r;
}
\end{lstlisting}
\end{minipage}

&

\begin{minipage}{1.5in}
\begin{lstlisting}
$\textbf{count}$(n,a){

  while ($\fbox{\parbox[c][0.5em][c]{0.25\textwidth}{n > 0}}$){
    $\fbox{\parbox[c][3.1em][c]{0.85\textwidth}{
    t := n \% 10; \\
    a[t] := a[t] + 1; \\
    n := n / 10; }}$
  };
  return a;
  
}
\end{lstlisting}
\end{minipage}

\\

(a) Problem1 \quad

&

\quad\quad\quad(b) Problem 2

\\ 

\begin{minipage}{1.5in}
\begin{lstlisting}
$\textbf{sum}$(n){
  r := 0; 
  while ($\fbox{\parbox[c][0.5em][c]{0.25\textwidth}{n > 0}}$){
    $\fbox{\parbox[c][5.7em][c]{0.69\textwidth}{
    t := n; \\
    while (t > 0)\begin{scriptsize}\{\end{scriptsize} \\ 
    \begin{minipage}{1.0em}~\end{minipage} 
     r := r + t;  \\
     \begin{minipage}{1.0em}~\end{minipage} 
     t := t - 1;  \\
    \begin{scriptsize}\}\end{scriptsize};              \\   
    n := n - 1;  }}$
  }; 
  return r;
}
\end{lstlisting}
\end{minipage}

&

\begin{minipage}{1.5in}
\begin{lstlisting}
$\textbf{abssum}$(a, len){ 
  r := 0;
  i := 0;
  while (i < len){
    if ($\fbox{\parbox[c][0.5em][c]{0.40\textwidth}{a[i] < 0}}$) 
     {$\fbox{\parbox[c][0.5em][c]{0.70\textwidth}{r := r - a[i];}}$}
    else 
     {$\fbox{\parbox[c][0.5em][c]{0.70\textwidth}{r := r + a[i];}}$}
    i := i + 1;
  };
  return r;
}
\end{lstlisting}
\end{minipage}

\\ [0.5em]

(c) Problem 3 \quad

&

\quad\quad\quad (d) Problem 4 

\end{tabular}
\caption{Synthesized results by $\SIMPL$ (in the boxes)}
\label{fig:solution}
\end{figure}

\section{Showcase}
\label{sec:motive}
In this section, we showcase $\SIMPL$ with four programming problems
that most beginners feel difficult to solve.  To use $\SIMPL$,
students need to provide (1) a partial program, (2) a set of
input-output examples, and (3) resources that $\SIMPL$ can use.  The
resources consist of a set of integers, a set of integer-type
variables, and a set of array-type variables.  The goal of $\SIMPL$ is
to complete the partial program w.r.t. the input-output
examples, using only the given resources.

\vspace*{-3mm}
\paragraph{Problem 1 (Reversing integer)}
The first problem is to write a function that reverses a given
integer.  For example, given integer 12, the function should return
21.  Suppose a partial program is given as
\begin{lstlisting}
$\textbf{reverse}$ (n){ r := 0; while($\textbf{?}$){$\textbf{?}$}; return r;}
\end{lstlisting}
where \textbf{\texttt{?}} denotes holes that need to be completed.
Suppose further $\SIMPL$ is provided with
input-output examples $\myset{1 \mapsto 1,  12\mapsto 21, 123 \mapsto 321}$, 
integers $\myset{0,1,10}$, 
and integer variables $\myset{{\mbox{\tt n}},\mbox{\tt r},\mbox{\tt x}}$.

Given this problem, $\SIMPL$ produces the solution 
in Figure~\ref{fig:solution}(a) in 2.5 seconds. 
Note that, $\SIMPL$ finds out that the integer `1' is 
unnecessary and the final program does not contain it. 
Also, $\SIMPL$ does not require sophisticated examples,
so that $\SIMPL$ can be easily used by inexperienced students.


\vspace*{-3mm}
\paragraph{Problem 2 (Counting)}
\label{sec:count-freq}
The next problem is to write a function that counts the number of each
digit in an integer. 
The program takes an integer and an array as inputs, where each
element of the array is initially 0. As output, the program returns
that array but now each array element at index $i$ stores the number
of $i$s that occur in the given integer.  
For example, when a tuple $(220, \arr{0,0,0})$ is given, the
function should output $\arr{1,0,2}$; 0 occurs once, 1 does not occur,
and 2 occurs twice in `220'.
Suppose the partial program is given as
\begin{lstlisting}
$\textbf{count}$(n,a){ while($\textbf{?}$){$\textbf{?}$}; return a;}
\end{lstlisting}
with examples $\myset{(11,\arr{0,0})\mapsto\arr{0,2}, 
(220,\arr{0,0,0}) \mapsto \arr{1,0,2}}$, 
integers $\myset{0,1,10}$, 
integer variables $\myset{\mbox{\tt i},\mbox{\tt n},\mbox{\tt t} }$, 
and an array variable $\{\mbox{\tt a}\}$.

For this problem, $\SIMPL$ produces the program 
in Figure~\ref{fig:solution}(b) in 0.2 seconds.
Note that $\SIMPL$ uses a minimal set of resources; 
$\texttt{i}$ is not used though it is given as usable.

\vspace*{-3mm}
\paragraph{Problem 3 (Sum of sum)}
The third problem is to compute 
$1+(1+2)+...+(1+2+...+n)$ for a given integer $n$.
Suppose the partial program
\begin{lstlisting}
$\textbf{sum}$(n){ r := 0; while($\textbf{?}$){$\textbf{?}$}; return r;}
\end{lstlisting}
is given with examples 
$\myset{1 \mapsto 1, 2 \mapsto 4, 3 \mapsto 10, 4 \mapsto 20}$, 
integers $\myset{0,1}$, 
and integer-type variables 
$\myset{\mbox{\tt n},\mbox{\tt t}, \mbox{\tt r}}$.

Then, $\SIMPL$ produces the program 
in Figure~\ref{fig:solution}(c) in 37.6 seconds.
Note that $\SIMPL$ newly introduced a nested loop, which is absent in the
partial program.

\vspace*{-3mm}
\paragraph{Problem 4 (Absolute sum)}
The last problem is to sum the absolute values 
of all the elements in a given array.
We provide the partial program: 
\begin{lstlisting}
$\textbf{abssum}$(a, len){ r := 0; i := 0;
  while(i < len){ if($\textbf{?}$){$\textbf{?}$} else{$\textbf{?}$}; i:=i+1;}; 
return r;}
\end{lstlisting}
where the goal is to complete the condition and bodies of the
if-statement. 
Given a set of input-output examples 
$\myset{(\arr{-1,-2},2) \mapsto 3, (\arr{2,3,-4},3)\mapsto9}$, 
an integer $\myset{0}$, integer variables $\{\mbox{\tt r},\mbox{\tt i}\}$,
and an array variable $\{\mbox{\tt a} \}$,
$\SIMPL$ produces the program in Figure~\ref{fig:solution}(d) in 12.1 seconds.




 


\section{Problem Definition}
\label{sec:language}
{\bf Language}
We designed an imperative language that is small yet expressive enough to 
deal with various programming problems in introductory courses.
The syntax of the language is defined by the following grammar:
\[
\begin{small}
\begin{array}{l}
  \abinop \to + \mid - \mid * \mid / \mid \%,\quad\bbinop \to\; = \mid > \mid < \\ [0.2em]
  
  l \to  x \mid x[y],  
  \quad
  a  \to n \mid l \mid l_1 \abinop l_2 \mid l \abinop n \mid~ \ahole \\ [0.2em]

  b \to {\true} \mid {\false} \mid  l_1 \bbinop l_2 \mid l \bbinop n \mid 
        b_1 \myand b_2 \mid b_1 \myor b_2 \mid \mynot b \mid~ \bhole \\ [0.2em]

  c \to  l := a \mid \skipcmd \mid c_1;c_2 \mid 
        \ifcmd~b~c_1~c_2 \mid \whilecmd~b~c \mid ~\chole 
\end{array}
\end{small}
\]
An l-value ($l$) is a variable ($x$) or an array reference ($x[y]$).
An arithmetic expression ($a$) is an integer constant $(n)$, an
l-value ($l$), or a binary operation ($\abinop$). 
A boolean
expression ($b$) is a boolean constant (${\true}, \false$), 
a binary relation ($\bbinop$), a negation ($\mynot b$), 
or a logical conjunction ($\myand$) and disjunction ($\myor$).  
Commands include assignment ($l := a$), 
skip ($\skipcmd$), sequence ($c_1;c_2$), 
conditional statement ($\ifcmd~b~c_1~c_2$), and while-loop ($\whilecmd~b~c$).

A program $P = (x,c,y)$ is a command with input and output variables,
where $x$ is the input variable, $c$ is the command, and $y$ is the
output variable. The input and output variables $x$ and $y$ can be
either of integer or array types. For presentation brevity, we assume
that the program takes a single input, but our implementation 
supports multiple input variables as well. 


An unusual feature of the language is that it allows to write incomplete
programs. Whenever uncertain, any arithmetic expressions, boolean
expressions, and commands can be left out with holes
($\ahole, \bhole, \chole$).  The goal of our synthesis algorithm is to
automatically complete such partial programs.

The semantics of the language is defined for programs without holes. 
Let $\Var$ be the set of program variables, which is partitioned into
integer and array types, i.e., $\Var  = \iVar \uplus \aVar$. 
A memory state 
\[
m \in \mbm = \Var \to \mbv, \quad v \in \mbv = \mbz + \mbz^* 
\] 
is a partial function from variables to values ($\mbv$).
A value is either an integer or an array of integers. 
An array $a \in \mbz^*$ is a sequence of integers. For instance, we write
$\arr{1,2,3}$ for the array of integers 1, 2, and 3. We write $|a|$, $a_i$, and $a_i^k$ for the length of $a$, the
element at index $i$, and the array $a_0\dots a_{i-1} k a_{i+1}
\dots a_{|a|-1}$, respectively. 

The semantics of the language is defined by the functions:  
\[\small
\sema{a}: \mbm \to \mbv, \quad \semb{b}: \mbm \to \mbb, \quad
\semc{c}: \mbm \to \mbm
\]
where $\sema{a}$, $\semb{b}$, and $\semc{c}$ denote the semantics of
arithmetic expressions, boolean expressions, and commands,
respectively. Figure~\ref{fig:semantics} presents the denotational
semantics, where $\fix$ is a fixed point operator. 
Note that the semantics for holes is undefined.

\begin{figure}
\[
\small
\begin{array}{r@{~}c@{~}l}
\sema{n}(m) &=& n \\
 \sema{x}(m) &=& m(x) \\
\sema{x[y]}(m) &=& m(x)_{m(y)} \\
\sema{l_1 \abinop l_2}(m) &=& \sema{l_1}(m) \abinop \sema{l_2}(m)\\
\sema{l \abinop n}(m) &=& \sema{l}(m) \abinop n \\
\end{array}
\]
\[
\small
\begin{array}{r@{~}c@{~}l}
\semb{\true}(m) &=& \true \\
\semb{\false}(m) &=& \false \\
\semb{l_1 \bbinop l_2}(m) &=& \sema{l_1}(m) \bbinop \sema{l_2}(m) \\
\semb{l \bbinop n}(m) &=& \sema{l}(m) \bbinop n \\
\semb{b_1 \land b_2}(m) &=& \semb{b_1}(m) \land \semb{b_2}(m) \\
\semb{b_1 \vee b_2}(m) &=& \semb{b_1}(m) \vee \semb{b_2}(m) \\
\semb{\neg b}(m) &=& \neg \semb{b}(m)
\end{array}
\]
\[
\small
\begin{array}{r@{~}c@{~}l}
\semc{x:=a}(m) &=& m[x \mapsto \sema{a}(m)] \\
\semc{x[y]:=a}(m) &=& m[x \mapsto m(x)_{m(y)}^{\sema{a}(m)}] \\
\semc{\skipcmd}(m) &=& m \\
\semc{c_1;c_2}(m) &=& (\semc{c_2} \circ \semc{c_1})(m) \\
\semc{\ifcmd~b~c_1~c_2}(m) &=& \cond(\semb{b}, \semc{c_1}, \semc{c_2})(m) \\
\semc{\whilecmd~b~c}(m) &=& (\fix~F)(m) \\
\mbox{where}~F(g) &=& \cond(\semb{b}, g \circ \semc{c}, \lambda x.x)
  \\
\cond(p,f,g)(m) &=& \left\{
\begin{array}{ll}
f(m) & \mbox{if}~p(m) = \true \\
g(m) & \mbox{if}~p(m) = \false
\end{array}
\right.
\end{array}
\]
\caption{Semantics of the language}
\label{fig:semantics}
\end{figure}


\vspace*{1mm}
\noindent
{\bf Synthesis Problem}
A synthesis task is defined by the five components: 
\[
((x,c,y), \examples, \integers, \iVar, \aVar)
\]
where $(x,c,y)$ is an incomplete program with holes,
$\examples \subseteq \mbv \times \mbv$ is a set of input-output examples. 
$\integers \subseteq \mbz$ is a set of integers, $\iVar$ is a set of
integer-type variables, and $\aVar$ is a set of array-type variables.
The goal of our synthesis algorithm is to produce a complete command
$c$ without holes such that
\begin{itemize}
\item $c$ uses constants and variables in $\integers$ and $\iVar \cup \aVar$, and
\item $c$ is consistent with every input-output example:
\[
\forall (v_i,v_o) \in \examples.\; \big(\semc{c}([x \mapsto v_i])\big)(y) = v_o.
\]
\end{itemize}

\section{Synthesis Algorithm}
\label{sec:basic}
In this section, we present our synthesis algorithm that combines 
enumerative search with static analysis. 

\subsection{Synthesis as State-Search}
We first reduce the synthesis task into a state-search problem. 
Consider a synthesis task $((x,c,y), \examples, \integers, \iVar, \aVar)$. 
The corresponding search problem is defined by the transition
system $(\states, \to, \initialstate, \finalstates)$
where $\states$ is a set of states, $(\to) \subseteq \states \times \states$ is a transition relation,
$\initialstate \in \states$ is an initial state, and $\finalstates
\subseteq \states$ is a set of solution states.

\begin{itemize}
\item \textbf{States} : A state $s \in \states$ is a command possibly with holes, 
which is defined by the grammar in Section \ref{sec:language}. 

\item \textbf{Initial state} : An initial state $\initialstate$ is a partial command $c_0$.

\item \textbf{Transition relation} : Transition relation $(\to) \subseteq \states \times \states$ 
determines the state that is immediately reachable from a 
state. The relation is defined as a set of inference rules in Figure
\ref{fig:transition}. 
Intuitively, a hole can be replaced by an arbitrary expression 
(or command) of the same type. 
Given a state $s$, we write $\nextstates(s)$ for the set of all immediate next states, 
i.e., $\nextstates (s) = \{s' \mid s \to s'\}$.
We write $s\not\to$ for terminal states, i.e., states with no holes. 

\item \textbf{Solution states} : A state $s$ is a solution 
iff $s$ is a terminal state and it is consistent with all input-output examples:
\[
\begin{array}{l}
\solution (s) \iff \\
\qquad s \not\to\myland \forall (v_i,v_o) \in \examples.\; \big(\semc{s}([x \mapsto
  v_i])\big)(y) = v_o.
\end{array}
\]
\end{itemize}

\begin{figure*}
\[
\begin{footnotesize}
\begin{array}{c}

\infer{l:=a \to l:=a'}{a \to_a a'}  ~~
\infer{c_1;c_2 \to c_1';c_2}{c_1 \to c_1'} ~~
\infer{c_1;c_2 \to c_1;c_2'}{c_2 \to c_2'} ~~
\infer{\ifcmd~b~c_1~c_2 \to \ifcmd~b'~c_1~c_2}{b \to_b b'} ~~
\infer{\ifcmd~b~c_1~c_2 \to \ifcmd~b~c_1'~c_2}{c_1 \to c_1'} ~~
\infer{\ifcmd~b~c_1~c_2 \to \ifcmd~b~c_1~c_2'}{c_2 \to c_2'}  ~~
\infer{\neg b \to_b \neg b'}{b \to_b b'}  \\ [0.3em]

\infer{\whilecmd~b~c \to \whilecmd~b'~c}{b \to_b b'} ~~
\infer{\whilecmd~b~c \to \whilecmd~b~c'}{c \to c'} ~~
\infer{b_1 \land b_2 \to_b b_1' \land b_2}{b_1 \to_b b_1'} ~~
\infer{b_1 \land b_2 \to_b b_1 \land b_2'}{b_2 \to_b b_2'} ~~
\infer{b_1 \vee b_2 \to_b b_1' \vee b_2}{b_1 \to_b b_1'} ~~
\infer{b_1 \vee b_2 \to_b b_1 \vee b_2'}{b_2 \to_b b_2'} \\ [0.5em]

\infer{\chole \to  l:=\ahole}     {} \quad 
\infer{\chole \to \skipcmd}     {} \quad 
\infer{\chole \to \chole;\chole} {} \quad \infer{\chole \to
\ifcmd~\bhole~\chole~\chole}{} \quad
\infer{\chole \to \whilecmd~\bhole~\chole}{} \quad
\infer{\bhole \to_b true}     {} \quad 
\infer{\bhole \to_b false}     {} \quad
\infer[]{\bhole \to_b l_1 \bbinop l_2}  {}  \\ [0.5em]
\infer[]{\bhole \to_b l \bbinop n}     {} \quad

\infer{\bhole \to_b \bhole \land \bhole}     {} \quad 
\infer{\bhole \to_b \bhole \vee \bhole}     {} \quad 
\infer{\bhole \to_b \neg\bhole}     {} \quad 
\infer[]{\ahole \to_a n}{} \quad
\infer[]{\ahole \to_a l}{} \quad
\infer[]{\ahole \to_a l_1 \abinop l_2 }{} \quad
\infer[]{\ahole \to_a l \abinop n }{} 
\end{array}
\end{footnotesize}
\]
\caption{Transition Relation $(n \in \integers$, $l \in \iVar \cup \{x[y] \mid x \in \aVar \land y \in \iVar\})$   }
\label{fig:transition}
\end{figure*}

\begin{algorithm}[t]
\caption{Synthesis Algorithm}
\label{algpseudo}
\begin{algorithmic}[1]
\Require A synthesis problem $((x,c_0,r),\examples,\integers,\iVar,\aVar)$
\Ensure A complete program consistent with $\examples$
\State $W \gets \{ c_0 \} $
\Repeat
  \State Pick the smallest state $s$ from $W$
  \If{$s$ is a terminal state}  
\State {\bf if} $\solution(s)$ {\bf then} \Return $s$
  \Else 
   \State {\bf if}~$\neg\prune(s)$~{\bf then}~$W \gets W \union \nextstates(s)$  
  \EndIf
\Until{$W \neq \emptyset$}
\end{algorithmic}
\end{algorithm}

\subsection{Baseline Search Algorithm} 
\label{sec:baseline}

Algorithm~\ref{algpseudo} shows the basic architecture of our
enumerative search algorithm.  The algorithm initializes the
workset $W$ with $c_0$ (line 1).  Then, it picks
a state $s$ with the smallest size and removes the state from the workset (line 3).
If $s$ is a solution state, the algorithm terminates and $s$ is
returned (line 5).   For a non-terminal state,
the algorithm attempts to prune the state by invoking the function
$\prune$ (line
7). 
If pruning fails, the
next states of $s$ are added into the workset and the loop
repeats. The details of our pruning technique is described
in Section \ref{sec:static-analysis-pruning}. At the moment, assume $\prune$ always fails. 

The baseline algorithm implicitly performs two well-known optimization
techniques. First, it maintains previously explored states and never
reconsider them. Second, more importantly, it normalizes states so that
semantically-equivalent programs are also syntactically the same. For
instance, suppose $(r:=0; r:=x*0; \chole)$ is the current
state. Before pushing it to the workset, we first normalize it to
$(r:=0; \chole)$.  To do so, we use four code optimization techniques:
constant propagation, copy propagation, dead code elimination, and
expression simplification~\cite{dragonbook}.  These two techniques
significantly improve the speed of enumerative search.

In addition, the algorithm considers terminating programs only. 
Our language has unrestricted loops, so the basic algorithm may synthesize non-terminating programs.
To exclude them from the search space, we use syntactic heuristics to
detect potentially non-terminating loops.  The heuristics are: 1) we
only allow boolean expressions of the form $x < y$ (or $x > n$) in
loop conditions, 2) the last statement of the loop body must increase
(or decrease) the induction variable $x$, and 3) $x$ and $y$ are not
defined elsewhere in the loop. 




\subsection{Pruning with Static
  Analysis}\label{sec:static-analysis-pruning}
Now we present the main contribution of this paper, 
pruning with
static analysis. Static analysis allows to safely identify states that
eventually fail to be a solution.  We first define the notion of
failure states.
\begin{definition}
A state $s$ is a failure state, denoted $\dead(s)$,
iff every terminal state $s'$ reachable from $s$ is not a solution, i.e.,
\[
\dead(s) \iff 
((s \to^* s') \myand s' \not\to \implies \neg \solution(s')).
\]
\end{definition}
Our goal is to detect as many failure states as possible. 
We observed two typical cases of failure states 
that often show up during the baseline search algorithm.

\begin{example}
\label{motivex:itv}
Consider the program in Figure~\ref{fig:pruning}(a) and input-output
example ${(1,1)}$. 
When the program is executed with $n=1$, no matter how the hole $(\ahole)$ gets instantiated,
the output value $r$ is no less than 2 at the return statement.
Therefore,  the program cannot but fail to satisfy the example $(1,1)$.
\end{example}

\begin{example}
\label{motivex:sym}
Consider the program in Figure~\ref{fig:pruning}(b) and input-output
example ${(1,1)}$. 
Here, we do not know the exact values of $x$ and $r$, but we know that
$10*x = 1$ must hold at the end of the program. 
However, there exists no such integer $x$, and we conclude the
partial program is a failure state. 
\end{example}

\lstset{fancyvrb=true,basicstyle=\ttfamily\footnotesize,columns=flexible,mathescape,numbers=none,numberstyle=\tiny}

\begin{figure}[t]
\begin{center}
\begin{tabular}{c@{\quad\quad}c@{\quad\quad}r}
\begin{minipage}{1.5in}
\begin{lstlisting}
  example1(n){  
    r := 0;
    while (n > 0){
      r := n + 1;
      n := $\ahole$;
    };
    return r;    
  }
\end{lstlisting}
\end{minipage}
&
\begin{minipage}{1.5in}
\begin{lstlisting}
  example2(n) {
    r := 0; 
    while (n > 0){
           $\chole$;
      r := x * 10;
      n := n / 10;
    }; 
    return r;
  }
\end{lstlisting}
\end{minipage}
\\
(a) \quad\quad
& (b) \quad
\end{tabular}
\end{center} 
\caption{States that are pruned away}
\label{fig:pruning}
\end{figure}

\noindent
{\bf Static Analysis}
We designed a static analysis that aims to effectively identify
these two types of failure states.
To do so, our analysis combines numeric and 
symbolic analyses; the numeric
analysis is designed to detect the cases of Example~\ref{motivex:itv} and
the symbolic analysis for the cases of Example~\ref{motivex:sym}.
The abstract domain of the analysis is defined as follows:
\[
\amem \in \aMem = \Var \to \aValue, \quad
\avalue \in \aValue = \mbi \times \mbs  
\]
An abstract memory state $\amem$ maps variables to
abstract values ($\aValue$). An abstract value is a pair of intervals
($\mbi$) and symbolic values ($\mbs$). 
The domain of intervals is standard~\cite{aiframework}: 
\[
\mbi = (\{\bot\} \cup \{[l, u] \mid l, u \in \mbz \cup \{-\infinity,
+\infinity\} \land l \le u \}, \sqsubseteq_\mbi).
 \]
For symbolic analysis, we define the following flat domain: 
\[
\mbs = (\SE_{\bot}^{\top}, \sqsubseteq_\mbs)~~\mbox{\it where}~~
\SE \to n \mid \sym_x~(x \in \Var_i) \mid \SE \abinop \SE
\]
A symbolic expression $\se \in \SE$ is a constant ($n$), a
symbol ($\sym_x$), or a binary operation with symbolic expressions. We
introduce symbols one for each integer-type variable in the program. The symbolic
domain is flat and has the partial order:
$s_1
\sqsubseteq_\mbs s_2\iff (s_1 = \bot) \vee (s_1 = s_2) \vee (s_2 =
\top)$. 
We define the abstraction function 
$\alpha: \mbv \to \aValue$ that transforms concrete values to abstract values: 
\[
\begin{array}{r@{\!~~}c@{\!~~}l}
\alpha(n) &=& ([n,n],n) \\
 \alpha(n_1\dots n_k) &=& ([\min
\myset{n_1,\dots,n_k}, \max \myset{n_1,\dots,n_k}], \top).
\end{array}
\]
The abstract semantics is defined in Figure~\ref{fig:abstract-semantics} by the functions:
\[
\abssema{a}:\aMem \to \aValue, \quad
\abssemb{b}:\aMem \to \aBool, \quad
\abssemc{c}:\aMem \to \aMem
\]
where $\aBool = \myset{\atrue,\afalse}_\bot^\top$ is the abstract
boolean lattice. 

\begin{figure}
\begin{footnotesize}
\[
\begin{array}{rcl}
\abssema{n}(\amem) &=& ([n,n],n) \\
\abssema{l}(\amem) &=& \amem(x)~(l = x~\mbox{or}~x[y]) \\
\abssema{l_1 \abinop l_2}(\amem) &=& \abssema{l_1}(\amem) {\aabinop} \abssema{l_2}(\amem) \\
\abssema{l \abinop n}(\amem) &=& \abssema{l_1}(\amem) \aabinop n \\
\abssema{\ahole}(\amem) &=& ([-\infty,+\infty],\top) \\[0.3em]
\abssemb{\true}(\amem) &=& \atrue~ (\abssemb{\false}(\amem) = \afalse) \\
\abssemb{l_1 \bbinop l_2}(\amem) &=& \abssema{l_1}(\amem) \abbinop \abssema{l_2}(\amem) \\
\abssemb{l \bbinop n}(\amem) &=& \abssema{l}(\amem) \abbinop n \\
\abssemb{b_1 \land b_2}(\amem) &=& \abssemb{b_1}(\amem) \aland \abssemb{b_2}(\amem) \\
\abssemb{b_1 \vee b_2}(\amem) &=& \abssemb{b_1}(\amem) \avee \abssemb{b_2}(\amem) \\
\abssemb{\neg b}(\amem) &=& \aneg \abssemb{b}(\amem) \\
\abssemb{\bhole}(\amem) &=& \top \\[0.3em]
\abssemc{x:=\ahole}(\amem) &=& \amem[x \mapsto ([-\infty,+\infty],\sym_x)] \\
\abssemc{x[y]:=\ahole}(\amem) &=& \amem[x \mapsto ([-\infty,+\infty],\top)] \\
\abssemc{x:=a}(\amem) &=& \amem[x \mapsto \abssema{a}(\amem)] \\
\abssemc{x[y]:=a}(\amem) &=& \amem[x \mapsto \abssema{a}(\amem)\sqcup \amem(x)] \\
\abssemc{\skipcmd}(\amem) &=& \amem \\
\abssemc{c_1;c_2}(\amem) &=& (\abssemc{c_2} \circ \abssemc{c_1})(\amem) \\
\abssemc{\ifcmd~b~c_1~c_2}(\amem) &=& \acond(\abssemb{b}, \abssemc{c_1}, \abssemc{c_2})(\amem) \\
\abssemc{\whilecmd~b~c}(\amem) &=& (\afix~\aF)(\amem) \\
\mbox{where}~\aF(g) &=& \acond(\abssemb{b}, g \circ \abssemc{c},
                        \lambda x.x) \\
\abssemc{\chole}(\amem)(x) &=& 
\left\{
\begin{array}{ll}
([-\infty,+\infty],\sym_x) & x \in \iVar \\
([-\infty,+\infty],\top) & x \in \aVar \\
\end{array}
\right.
 \\ [0.3em]
\end{array}
\]
\end{footnotesize}
\caption{Abstract semantics}
\label{fig:abstract-semantics}
\end{figure}

Intuitively, the abstract semantics over-approximates
the concrete semantics of \textit{all} terminal states that are
reachable from the current state.
This is done by defining the sound semantics for holes:
$\abssema{\ahole}(\amem)$, $\abssemb{\bhole}(\amem)$, and
$\abssemc{\chole}({\amem})$. An exception is that
integer variables get assigned symbols, rather than $\top$, in order
to generate symbolic constraints on integer variables. 

In our analysis, array elements are abstracted into 
a single element. Hence, the definitions of 
$\abssema{x[y]}$
and $\abssemc{x[y]:=a}$
do not involve $y$. Because an abstract array cell may represent
multiple concrete cells, arrays are weakly updated by joining ($\join$) old and
new values.
For example, in memory state 
$\amem =[x \mapsto ([5,5], \top),...]$, 
$\abssemc{x[y]:=1}(\amem)$ 
evaluates to $[x \mapsto ([1,5], \top),...]$.

For while-loops, the analysis performs a sound fixed point
computation. 
If the computation does not reach a fixed point after a fixed number of iterations,
we apply widening for infinite interval domain, 
in order to guarantee the termination of the analysis.
We use the standard widening operator in~\cite{aiframework}.
The function $\afix$ and $\acond$ in Figure~\ref{fig:abstract-semantics} denote
a post-fixed point operator and a sound abstraction of $\cond$,
respectively. 

\vspace*{1mm}
\noindent
{\bf Pruning}
Next we describe how we do pruning with the static analysis.
Suppose we are given examples $\examples \subseteq \mbv \times \mbv$ and a state $s$ with input ($x$)
and output ($y$) variables. 
For each example $(v_i,v_o) \in \examples$, we first run the static
analysis with the input $\alpha(v_i)$ and obtain the analysis result $(itv_s,\se_s):$
\[
(itv_s,\se_s) = (\abssemc{s}([x \mapsto \alpha(v_i)])(y).
\]
We only consider the case when  $itv_s = [l_s,u_s]$ (when $itv_s=\bot$, the program
is semantically ill-formed and therefore we just prune out the state). 
Then, we obtain the interval abstraction $[l_o,u_o]$ of the output
$v_o$, i.e., $([l_o,u_o],-) = \alpha(v_o)$, and generate the
constraints $C_{(v_i,v_o)}^s$: 
\[
\begin{small}
\begin{array}{l}
C_{(v_i,v_o)}^s = (l_s \le l_o \land u_o \le u_s)\land (se \in\SE \implies l_o \le se \le u_o).
\end{array}
\end{small}
\] 
 
The first (resp., second) conjunct means that the interval (resp., symbolic) analysis result must
over-approximate the output example. 
We prune out a state $s$ iff $C_{(v_i,v_o)}^s$ is unsatisfiable for
some example $(v_i,v_o) \in \examples$:
\begin{definition} The predicate $\prune$ is defined as follows:
\[
\prune(s) \iff \mbox{$C_{(v_i,v_o)}^s$ is
unsatisfiable for some $(v_i,v_o) \in \examples$}.
\]
\end{definition}
The unsatisfiability can be easily checked, for instance, with an off-the-shelf SMT
solver. 
Our pruning is safe: 
\begin{theorem}[Safety]
\label{lem:pruning-sound}
$\forall s \in S.\; \prune(s) \implies \dead(s)$. 
\end{theorem} 
\noindent
That is, we prune out a state only when it is a failure
state, which formally guarantees that the search algorithm with our
pruning finds a solution if and only if the baseline algorithm
(Section \ref{sec:baseline}) does so. 



\myskip{
\newpage

\begin{figure}[t]
\[
\begin{footnotesize}
\begin{array}{rcl}
\abssema{a} &\in& \hat{\mbm} \to \hat{\mbv} \\
\abssema{n}(\hat{m}) &=& ([n,n], n)  \\
\abssema{l}(\hat{m}) &=&
                				\left\{
 						  \begin{array}{ll}
						    \hat{m}(x) & \textrm{if}~l=x\\ 
						    \hat{m}(x) & \textrm{if}~l=x[y]\\ 
						  \end{array}
					\right.   \\
\abssema{l_1\abinop l_2}(\hat{m}) &=& \abssema{l_1}(\hat{m})~\hat{\abinop}~\abssema{l_2}(\hat{m}) \\
\abssema{l \abinop n}(\hat{m}) &=& \abssema{l}(\hat{m})~\hat{\abinop}~\abssema{n}(\hat{m}) \\ 
\abssema{\ahole}(\hat{m}) &=& ([-\infinity,+\infinity],\top)   \\ [1.0em]

\abssemb{b} &\in& \hat{\mbm} \to \hat{\mbm} \\
\abssemb{\bhole}(\hat{m}) &=& \hat{m} \\
\abssemb{b} (\hat{m}) &=& \lub\{\hat{m}' \ple_\mbm \hat{m} \mid \hat{m}' \models b\} \\  [1.0em]

\abssemc{c} &\in& \hat{\mbm} \to \hat{\mbm} \\  
\abssemc{l:=\ahole}(\hat{m}) &=&
                				\left\{
 						  \begin{array}{ll}
						    \hat{m}[x \xmapsto[]{~~} \langle[-\infinity,+\infinity], \beta_x \rangle] & \textrm{if}~l=x\\ 
						    \hat{m}[x \xmapsto[]{w}  \langle[-\infinity,+\infinity], \top \rangle]  & \textrm{if}~l=x[y]\\ 
						  \end{array}
					\right.   \\
\abssemc{l:=a}(\hat{m}) &=&    
                				\left\{
 						  \begin{array}{ll}
						    \hat{m}[x \xmapsto[]{~~} \abssema{a}(\hat{m})] & \textrm{if}~l=x\\ 
						    \hat{m}[x \xmapsto[]{w} \abssema{a}(\hat{m})]  & \textrm{if}~l=x[y]\\ 
						  \end{array}
					\right.   \\
\abssemc{\skipcmd}(\hat{m}) &=& \hat{m} \\
\abssemc{c_1;c_2}(\hat{m}) &=& \abssemc{c_2} \circ \abssemc{c_1} (\hat{m}) \\
\abssemc{\ifcmd~b~c_1~c_2}(\hat{m}) &=& \cond (b,c_1,c_2) (\hat{m}) \\
\abssemc{\whilecmd~b~c}(\hat{m}) &=&  \abssemb{\neg b}(\fix~F)(\hat{m}) \\
                                                        &&
                                                           \textrm{where}~F=\cond(b, c; \whilecmd~b~c, \skipcmd)  \\   
\abssemc{\chole}(\hat{m}) &=& 
\lambda x \in \Var. \;                 				\left\{
 						  \begin{array}{ll}
						    \hat{m}[x
                                                    \xmapsto[]{~~}
                                                    \langle[-\infinity,+\infinity],
                                                    \beta_x \rangle] &
                                                                       \textrm{if}~x
                                                                       \in
                                                    \Var_i\\ 
						    \hat{m}[x
                                                    \xmapsto[]{w}
                                                    \langle[-\infinity,+\infinity],
                                                    \top \rangle]  &
                                                                     \textrm{if}~x
                                                                     \in
                                                    \Var_a\\ 
						  \end{array}
					\right.   \\
				                                                                                                             
\end{array}
\end{footnotesize}
\]
\[
\begin{footnotesize}
\begin{array}{l}
\cond (f,g,h) (\hat{m}) \\ =
\left\{
\begin{array}{ll}
\abssemc{g}(\abssemb{f}(\hat{m}))   & \hat{m} \models f \\           
\abssemc{h}(\abssemb{\neg f}(\hat{m}))  &\hat{m} \not\models f \\              
\abssemc{g}(\abssemb{f}(\hat{m})) \join~\abssemc{h}(\abssemb{\neg f}(\hat{m}))   &\textrm{otherwise}
\end{array}   
\right.
\end{array}
\end{footnotesize}
\]
\caption{Abstract Semantics}
\label{fig:abstract-semantics}
\end{figure}

\paragraph{Abstract interpretation based sound analysis}
\begin{figure}[t]
\[
\begin{footnotesize}
\begin{array}{rcl}
\alpha_\mbv  \in \power{\mbv} \to \hat{\mbv}
&
\alpha_\mbi   \in  \power{\mbv} \to \mbi
&
\alpha_\mbs   \in  \power{\mbv} \to \mbs 
\end{array}
\end{footnotesize}
\]
\[
\begin{footnotesize}
\begin{array}{rcl}
\alpha_\mbv (v) &=& (\alpha_\mbi (v), \alpha_\mbs (v)) \\ [0.5em]


\alpha_\mbi (v) &=& \left\{
 				\begin{array}{ll}
  				    \bot & \textrm{if}~v= \emptyset \\
				    $[$\mins(v),\maxs(v)$]$ &\textrm{otherwise}
				 \end{array}
				\right. \\ [1.0em]


\alpha_\mbs (v) &=& \left\{
 				\begin{array}{ll}
  				    \bot & \textrm{if}~v = \emptyset \\
				    (n/2=0)?\even:\odd  & \textrm{if}~v = \{n\}_{i\ge 1}   \\
				    \top &\textrm{otherwise}
				 \end{array}
				\right. \\ \\
\end{array}		
\end{footnotesize}					
\]
\[
\begin{footnotesize}
\begin{array}{lcr}
\gamma_\mbv  \in  \hat{\mbv} \to \power{\mbv} 
&
\gamma_\mbi \in \mbi \to \power{\mbv}
&
\gamma_\mbs \in \mbs \to \power{\mbv} 
\end{array}
\end{footnotesize}
\]
\[
\begin{footnotesize}
\begin{array}{rcl}
\gamma_\mbv (i,s) &=& \gamma_\mbi (i) \cup \gamma_\mbs(s)  \\ [0.5em]


\gamma_\mbi (i) &=& \left\{
 				\begin{array}{ll}
 				    \emptyset & \textrm{if}~i= \bot \\
				    \{n \mid a \le n \le b \}^+ &  \textrm{if}~i=$[$a,b$]$
				 \end{array}
				\right. \\ [1.0em]

\gamma_\mbs (s) &=& \left\{
 				\begin{array}{ll}
  				    \emptyset & \textrm{if}~s= \bot \\
				    \{ 2n \mid n \in \mbz \}^+ & \textrm{if}~s=\even \\				    
				    \{2n+1 \mid n \in \mbz \}^+ &  \textrm{if}~s=\odd \\
				    \mbv(=\mbz^+) & \textrm{if}~s= \top 
				 \end{array}
				\right. 
\end{array}
\end{footnotesize}
\]
\caption{Galois connection of concrete $(\power{\mbv})$ and abstract $(\hat{\mbv})$ values}  
\label{fig:galois}
\end{figure}

\begin{figure}[t]
\[
\begin{footnotesize}
\begin{array}{rcl}
\abssema{a} &\in& \hat{\mbm} \to \hat{\mbv} \\
\abssema{n}(\hat{m}) &=& ([n,n], (n/2=0)?\even:\odd)  \\
\abssema{l}(\hat{m}) &=&
                				\left\{
 						  \begin{array}{ll}
						    \hat{m}[x \mapsto ([-\infinity,+\infinity],\top)] & \textrm{if}~l=x\\ 
						    \hat{m}[x \mapsto ([-\infinity,+\infinity],\top)] & \textrm{if}~l=x[y]\\ 
						  \end{array}
					\right.   \\
\abssema{l_1\abinop l_2}(\hat{m}) &=& \abssema{l_1}(\hat{m})~\hat{\abinop}~\abssema{l_2}(\hat{m}) \\
\abssema{l \abinop n}(\hat{m}) &=& \abssema{l}(\hat{m})~\hat{\abinop}~\abssema{n}(\hat{m}) \\ 
\abssema{\ahole}(\hat{m}) &=& ([-\infinity,+\infinity],\top)   \\ [1.0em]

\abssemb{b} &\in& \hat{\mbm} \to \hat{\mbm} \\
\abssemb{\bhole}(\hat{m}) &=& \hat{m} \\
\abssemb{b} (\hat{m}) &=& \lub\{\hat{m}' \ple_\mbm \hat{m} \mid \hat{m}' \models b\} \\  [1.0em]

\abssemc{c} &\in& \hat{\mbm} \to \hat{\mbm} \\  
\abssemc{l:=a}(\hat{m}) &=&    
                				\left\{
 						  \begin{array}{ll}
						    \hat{m}[x \xmapsto[]{~~} \abssema{a}(\hat{m})] & \textrm{if}~l=x\\ 
						    \hat{m}[x \xmapsto[]{w} \abssema{a}(\hat{m})]  & \textrm{if}~l=x[y]\\ 
						  \end{array}
					\right.   \\
\abssemc{\skipcmd}(\hat{m}) &=& \hat{m} \\
\abssemc{c_1;c_2}(\hat{m}) &=& \abssemc{c_2} \circ \abssemc{c_1} (\hat{m}) \\
\abssemc{\ifcmd~b~c_1~c_2}(\hat{m}) &=& \cond (b,c_1,c_2) (\hat{m}) \\
\abssemc{\whilecmd~b~c}(\hat{m}) &=&  \abssemb{\neg b}(\fix~F) \\
                                                        && \textrm{where}~F=\cond(b, c, \skipcmd)  \\   
\abssemc{\chole}(\hat{m}) &=& \forall x \in \vars. \hat{m}[x \mapsto ([-\infinity,+\infinity], \top)]				                                                                                                             
\end{array}
\end{footnotesize}
\]
\[
\begin{footnotesize}
\begin{array}{l}
\cond (f,g,h) (\hat{m}) \\ =
\left\{
\begin{array}{ll}
\abssemc{g}(\abssemb{f}(\hat{m}))   & \hat{m} \models f \\           
\abssemc{h}(\abssemb{\neg f}(\hat{m}))  &\hat{m} \not\models f \\              
\abssemc{g}(\abssemb{f}(\hat{m})) \join~\abssemc{h}(\abssemb{\neg f}(\hat{m}))   &\textrm{otherwise}
\end{array}   
\right.
\end{array}
\end{footnotesize}
\]
\caption{Abstract Semantic Functions}
\label{fig:semantics}
\end{figure}

We formalize our numerical analysis in the abstract interpretation framework~\cite{aiframework}.
Relying on the framework, the analysis guarantees soundness, i.e., 
the analyzed results consider all possible actual program behaviors.

A key of the analysis is to track actual (concrete) values
with abstract values, which are interval values $(\mbi)$ 
and parity values $(\mbs)$ respectively.

$\mbi$ is the interval lattice
which computes lower and upper bounds of program variables
with order relation $\sqsubseteq_\mbi$:
\[
\begin{footnotesize}
\mbi = \{\bot\} \cup \{[l, u] \mid l, u \in \mbz \cup \{-\infinity, +\infinity\} \land l \le u \} 
\end{footnotesize}
\]
\[
\begin{scriptsize}
\begin{array}{l}
i_1 \sqsubseteq_\mbi i_2  \iff 
                					\left\{
 						  \begin{array}{l}
						    i_1 = \bot ~ \myor \\
  						    i_2 = [-\infinity, +\infinity] ~\myor \\
						    (i_1 = [l_1,u_1]\myland i_2 = [l_2,u_2]\myland l_1 \ge l_2 \myland u_1 \le u_2).
						  \end{array}
						\right. \\												
\end{array}
\end{scriptsize}
\]
Parity lattice $\mbs =\{\bot, \even, \odd, \top\}$ determines
whether actual values are either even or odd
with ordering $(\bot \sqsubseteq_\mbs \even \sqsubseteq_\mbs\top)$
and $(\bot \sqsubseteq_\mbs \odd \sqsubseteq_\mbs \top)$.

To define the Galois connection
between the concrete and abstract domain, 
the concrete value domain $\mbv$ is lifted into $\power{\mbv}$,
where each value is represented as a singleton set,
and the domain of the concrete memory state is pointwisely lifted,
i.e., $\mbm = \Var \to \power{\mbv}$. 
Then, we have the following Galois connection:
\[
\begin{footnotesize}
\begin{array}{l}
\power{\mbv} \galois{\alpha_{\mbv}}{\gamma_{\mbv}} \hat{\mbv} ,
\quad
\Var \to \power{\mbv}   \galois{\alpha_{\mbm}}{\gamma_{\mbm}} \Var \to \hat{\mbv}
\end{array}
\end{footnotesize}
\]
where $\alpha_\mbm$ and $\gamma_\mbm$ are pointwise liftings of 
abstract function $\alpha_\mbv$ and concretization function $\gamma_\mbv$.
Figure~\ref{fig:galois} shows definitions of 
$\alpha_\mbv$ and $\gamma_\mbv$ 
with sub-abstraction functions for intervals $(\alpha_\mbi)$
and parity values $(\alpha_\mbs)$, 
accompanied by sub-concretization functions 
for intervals $(\gamma_\mbi)$ and parity values $(\gamma_\mbs)$.

An abstract memory state $\hat{m} \in \hat{\mbm} = \Var \to \hat{\mbv}$ 
is a finite map from variables to abstract values $(\hat{\mbv})$,
where the abstract values consist of the lattice of intervals 
and parity lattice, i.e., $\hat{\mbv} =\mbi \times \mbs$
with componentwise ordering:
\[
\begin{footnotesize}
(i_1,s_1) \sqsubseteq_\mbv (i_2,s_2) \iff i_1 \sqsubseteq_\mbi i_2  \land  s_1 \sqsubseteq_\mbs s_2.
\end{footnotesize}
\]

Note that we use the parity lattice 
as a finite symbolic expression domain, 
where $\even$ and $\odd$ represent 
symbolic expression $2\beta$ and $2\beta+1$ respectively. 
In a domain of infinite symbolic expressions,
the ordering is not always possible 
because non-linear integer arithmetic is undecidable.

\paragraph{Abstract semantic function}
Figure~\ref{fig:semantics} shows the abstract 
semantic functions that we use in the numerical analysis.

$\abssema{a}(\hat{m})$ computes
the abstract value for $a$ under the $\hat{m}$.
Arithmetic expressions, which involve
abstract binary operators $(\hat{\abinop})$,  
are inductively evaluated.
The analysis uses single abstraction
to represent all array elements for each array, 
hence the definition of $\abssema{x[y]}(\hat{m})$ 
do not involve $y$.
For the arithmetic hole $(\ahole)$, the analysis 
over-approximates the semantics as top value of each domain. 

$\abssemb{b}(\hat{m})$ confines
abstract values of variables which occur in $b$
so that the condition $b$ holds.
$\ple_\mbm$ means pointwise ordering for $\ple_\mbv$.
We write $\hat{m} \models b$ iff $b$ is true 
under $\hat{m}$, and $\hat{m} \not\models b$ 
means that $b$ is false under $\hat{m}$.
These conditions can checked 
by comparing interval values.
For soundness, no values are trimmed 
when the condition is undetermined $(\bhole)$.

$\abssemc{c}(\hat{m})$ possibly changes the abstract memory state 
by executing commands $c$ under $\hat{m}$. 
Similar to the case of $\abssema{x[y]}(\hat{m})$, 
array values are weakly updated $(\xmapsto[]{w})$, 
i.e., the array values are accumulated as single abstraction.
$\cond(f,g,h)(\hat{m})$ 
executes either $g$ or $h$ 
if $f$ evaluates to true or false under $\hat{m}$ correspondingly . 
If the truth value is unknown, the analysis performs 
pointwise join $(\join)$ after executing both $f$ and $g$ 
with confined memory states.
For a loop, the analysis computes 
the fixed point of the memory state
and returns the memory state confined by $\neg b$.
If the analysis does not reaches to the fixed point 
after some number of iterations, 
the analysis applies the same widening operator in~\cite{widening}
for the infinite interval domain 
to ensure the termination of the analysis. 
When command hole $(\chole)$ is encountered, 
the values of all the variables in $\vars$ are 
over-approximated, i.e., get top value. 

Let $s$ be a state.
We conclude the state $s$ is hopeless
iff the analyzed result of $s$ does not precede 
the abstracted output example value: 
\[
\begin{array}{l}
\hopeless (s) \iff \\
\qquad \exists (i,o) \in \examples. \alpha_\mbv(\{o\}) \not\ple_\mbv (\abssemc{s} \circ \alpha_\mbm ([x \mapsto \{i\}]))(r). 
\end{array}
\] 

Given a state $s$,
$\abssemc{s}$ is the over-approximations 
for all possible terminal states that are reachable from $s$.
Thus, we have the following lemma:
\begin{lemma}
\label{lem:over-approx}
For any state s, 
\[
 \underset{s \to^* s' \land s' \not\to}{\bigsqcup} \alpha_\mbm \circ \semc{s'} \ple \abssemc{s} \circ \alpha_\mbm
\]
where $\ple$ is the pointwise liftings of $\ple_\mbm$.
\end{lemma} 

By Lemma~\ref{lem:over-approx},
we have the following lemma which says 
our numerical analysis prunes the search space
in safe ways:
\begin{lemma}
\label{lem:pruning-sound}
For any state s,
\[
\hopeless(s) \implies \dead(s).
\]
\end{lemma} 
\begin{proof}
Suppose $\hopeless(s)$ holds.
From Lemma~\ref{lem:over-approx},
we obtain
\[
\exists (i,o) \in \examples. \alpha_\mbv(\{o\}) \not\ple_\mbv \underset{s \to^* s' \land s' \not\to}{\bigsqcup} (\alpha_\mbm \circ \semc{s'} ([x \mapsto \{i\}]))(r)
\]
which implies 
\[
((s \to^* s') \myand s'  \implies \forall (i,o) \in \examples. o \not= \; \big(\semc{s'}([x \mapsto i])\big)(y) 
\]
, i.e., $\dead(s)$.
\end{proof}

By Lemma~\ref{lem:pruning-sound}, our algorithm
guarantees to find a solution, if any.

}


\section{Evaluation}
\renewcommand{\arraystretch}{1.08}
\begin{table*}[t]
\begin{footnotesize}
\centering
\begin{tabular}{|c|c|l|r|r|r|r|r|r|r|}
\hline
\multicolumn{1}{|l|}{\multirow{2}{*}{Domain}} & \multirow{2}{*}{No} & \multicolumn{1}{c|}{\multirow{2}{*}{Description}}                 & \multicolumn{2}{c|}{Vars}                     & \multicolumn{1}{c|}{\multirow{2}{*}{Ints}} & \multicolumn{1}{c|}{\multirow{2}{*}{Exs}} & \multicolumn{3}{c|}{Time (sec)}  \\ \cline{4-5} \cline{8-10}
\multicolumn{1}{|l|}{}                        &                     & \multicolumn{1}{c|}{}                                             & \multicolumn{1}{c|}{\begin{scriptsize}IVars\end{scriptsize}} & \multicolumn{1}{c|}{\begin{scriptsize}AVars\end{scriptsize}} & \multicolumn{1}{c|}{}    & \multicolumn{1}{c|}{}   & \multicolumn{1}{c|}{Base}  & \multicolumn{1}{c|}{Base+Opt}  & \multicolumn{1}{c|}{Ours}            \\ \hline
\multirow{16}{*}{Integer}                    
					    & 1                   & Given $n$, return  $n!$.  													  & 2                     & 0                     & 2                                          & 4                                         			   & 0.0                         & 0.0                           & 0.0  \\  
                                              & 2                   & Given $n$, return  $n!!$ (i.e., double factorial).              					        		  & 3                     & 0                     & 3                                          & 4                              		                    & 0.0                         & 0.0                           &  0.0 \\      
                                              & 3                   & Given $n$, return  \begin{scriptsize}$\sum_{i=1}^{n} i$\end{scriptsize}.             		          & 3                     & 0                     & 2                                          & 4                                    		            & 0.1                         & 0.0                           &  0.0 \\       
                                              & 4                   & Given $n$, return  \begin{scriptsize}$\sum_{i=1}^{n} i^2$\end{scriptsize}.   		        		  & 4                     & 0                     & 2                                          & 3                          			                    & 122.4                     & 18.1                         &  0.3 \\  
                                              & 5                   & Given $n$, return  \begin{scriptsize}$\prod_{i=1}^{n} i^2$\end{scriptsize}. 			 	  & 4                     & 0                     & 2                                          & 3                                           			   & 102.9                     & 13.6                         &  0.2 \\
                                              & 6                   & Given $a$ and $n$, return $a^n$.                									  & 4                     & 0                     & 2                                          & 4                                             		   & 0.7                         & 0.1                           &  0.1 \\
                                              & 7                   & Given $n$ and $m$, return \begin{scriptsize}$\sum_{i=n}^{m} i$\end{scriptsize}.     		  & 3                     & 0                     & 2                                          & 3                                            		   & 0.2                         & 0.0                           &  0.0 \\
                                              & 8                   & Given $n$ and $m$, return \begin{scriptsize}$\prod_{i=n}^{m} i$\end{scriptsize}.      		  & 3                     & 0                     & 2                                          & 3                                      		            & 0.2                        & 0.0                           &  0.1 \\
                                              & 9                   & Count the number of digit for an integer.                      						       		  & 3                     & 0                     & 3                                          & 3                                             		   & 0.0                        & 0.0                          &  0.0 \\
                                              & 10                  & Sum the digits of an integer.                                  						        		  & 3                     & 0                     & 3                                          & 4                                             		   & 5.2                        & 2.2                           &  1.3  \\
                                              & 11                  & Calculate product of digits of an intger.                         					        		  & 3                     & 0                     & 3                                          & 3                                            		   & 0.7                      & 2.3                           &  0.3 \\
                                              & 12                  & Count the number of binary digit of an integer.                   							  & 2                     & 0                     & 3                                          & 3                                  		                    & 0.0                        & 0.0                           & 0.0   \\
                                              & 13                  & Find the $n$th Fibonacci number.                         		   						  & 3                     & 0                     & 3                                          & 4                                              		   & 98.7                      & 13.9                         &   2.6  \\
                                              & 14                  & Given $n$, return \begin{scriptsize}$\sum_{i=1}^{n}(\sum_{m=1}^{i} m))$\end{scriptsize}.     & 3                     & 0                     & 2                                          & 4                                            		           & $\bot$                   & 324.9                       &  37.6   \\
                                              & 15                  & Given $n$, return \begin{scriptsize}$\prod_{i=1}^{n}(\prod_{m=1}^{i} m))$\end{scriptsize}.    & 3                     & 0                     & 2                                          & 4                                           			   & $\bot$                   & 316.6                      &  86.9   \\
                                              & 16                  & Reverse a given integer.                                          								  & 3                     & 0                     & 3                                          & 3                                          		           & $\bot$                   & 367.3		           & 2.5    \\ \hline
\multirow{16}{*}{Array}          & 17                  & Find the sum of all elements of an array.                       							  & 3                     & 1                     & 2                                          & 2                                              		  & 8.1                         & 3.6                          &  0.9       \\
                                              & 18                  & Find the product of all elements of an array.                    							  & 3                     & 1                     & 2                                          & 2                                              		  & 7.6                         & 3.9                          &  0.9       \\
                                              & 19                  & Sum two arrays of same length into one array.	            						          & 3                     & 2                     & 2                                          & 2                                              		  & 44.6                       & 29.9                        &  0.2         \\
                                              & 20                  & Multiply two arrays of same length into one array.         						          & 3                     & 2                     & 2                                          & 2                                              		  & 47.4                       & 26.4                        &  0.3         \\
                                              & 21                  & Cube each element of an array.                                    						          & 3                     & 1                     & 1                                          & 2                                              		  & 1283.3                   & 716.1                     &   13.0    \\
                                              & 22                  & Manipulate each element into 4th power.                   								  & 3                     & 1                     & 1                                          & 2                                              		  & 1265.8                   & 715.5                     &  13.0      \\
                                              & 23                  & Find a maximum element.                              									  & 3                     & 1                     & 2                                          & 2                                              		  & 0.9                         &  0.7                         & 0.4        \\
                                              & 24                  & Find a minimum element.          			                     							  & 3                     & 1                     & 2                                          & 2                                              		  & 0.8                         &  0.3                         &  0.1         \\
                                              & 25                  & Add 1 to each element.                                            								  & 2                     & 1                     & 1                                          & 3                                             		 & 0.3                          &  0.0                 	   &  0.0       \\
                                              & 26                  & Find the sum of square of each element.                          							  & 3                     & 1                     & 2                                          & 2                                              		 & 2700.0                    &  186.2                     & 11.5       \\
                                              & 27                  & Find the multiplication of square of each element.            							  & 3                     & 1                     & 1                                          & 2                                              		 & 1709.8         		  &  1040.3               	   & 12.6      \\
                                              & 28                  & Sum the products of matching elements of two arrays.     							  & 3                     & 2                     & 1                                          & 3                                             		 & 20.5                        &  38.7                       &  1.5  \\
                                              & 29                  & Sum the absolute values of each element.                          							  & 2                     & 1                     & 1                                          & 2                                              		& 45.0                         &  50.5                       &  12.1    \\
                                              & 30                  & Count the number of each element.                             							  & 3                     & 1                     & 3                                          & 2                                              		& 238.9                       &   1094.1                  &  0.2   \\ \hline
					    & \multicolumn{5}{c}{Average}                                                                                &        &   $>$ 616.8             &  165.5       &  6.6    \\        
                                              \hline
\end{tabular}
\caption{Performance of $\SIMPL$. $\bot$ denotes timeout ($>$ 1 hour). Assume $\bot$ as 3,600 seconds for the average of ``Base".} 
\label{table}
\end{footnotesize}
\end{table*}

\noindent{\bf Experimental setup}
To evaluate our synthesis algorithm,
we gathered 30 introductory level problems from several online forums (Table~\ref{table}).\footnote{E.g., \url{http://www.codeforwin.in}}
The problems consist of tasks manipulating integers
and arrays. Some problems are non-trivial for novice students to solve; they require students to come up with various control structures
such as nested loops and combinations of
 loops and conditional statements.
The partial programs we used
are similar to those shown in Section~\ref{sec:motive};
they have one boolean expression hole $(\bhole)$,
and one or two command holes $(\chole)$. 
For each benchmark, 
we report the number of integer variables (IVars),
array variables (AVars),  
integer constants (Ints), 
and examples (Exs) provided, respectively.
All benchmark problems are publicly available with our tool.
Experiments were conducted on MacBook Pro 
with Intel Core i7 and 16GB of memory.

\noindent
{\bf Baseline Algorithm}
Table~\ref{table} shows the performance of our algorithm. 
The column ``Base" shows the running time
of our baseline algorithm that performs 
enumerative search without state normalization.
In that case, the average runtime was longer than 616 seconds,
and three of the benchmarks timed out ($>$ 1 hour). 
The column ``Base+Opt" reports the performance of the baseline with normalization. 
It shows that normalizing states 
succeeds to solve all benchmark problems and
improves the speed by more than 3.7 times on average, 
although it degrades the speed for some cases due to runtime normalization overhead. 

\noindent{\bf Pruning Effectiveness}
On top of ``Base+Opt'', we applied our static-analysis-guided pruning technique (the column ``Ours'').
The results show that our pruning technique is remarkably effective. 
It reduces the average time to 6.6 seconds, 
improving the speed of ``Base+Opt'' by 25 times. 
Note that $\SIMPL$ is able to synthesize the desired programs
from a few examples (Exs), requiring up to 4 examples. 



\section{Related Work}
\label{sec:related}
{\bf Computer-aided education}
Recently, program synthesis technology has revolutionized
computer-aided education. For instance, the technology has been used
in automatic problem
generation~\cite{probgen:aaai12,probgen:ijcai13,probgen:aaa14,probgen:ijcai15},
automatic grading~\cite{dfagrading:ijcai13}, and automatic solution
generation~\cite{solutiongen:pldi11}.

Our work is to use program synthesis for automated programming
education system. A large amount of work has been done to automate
programming
education~\cite{laura:1980,memo2:1981,lisptutor:1984,proust:1984,talus:1989,feedback:pldi13,feedback:fse14,feedback:fse16,apex},
which focuses primarily on providing feedback on students' programming submissions.
Our system, $\SIMPL$, has the following advantages over prior
works: 

\begin{itemize}
\item \textbf{Feedback on incomplete programs}: 
  Existing systems produce feedback only for complete programs;
  they cannot help students who do not know how to proceed
  further. In this case, $\SIMPL$ can help by automatically generating
  solutions starting from incomplete solutions. 

\item \textbf{No burden on instructor}:
Existing systems require instructor's manual effort. 
For example, the system in~\cite{feedback:pldi13} 
needs a correct implementation 
and a set of correction rules manually designed by the instructor. 
On the other hand, $\SIMPL$ does not require anything from the instructor.

An exception is~\cite{lisptutor:1984}, where an automatic LISP
feedback system is presented. However, the system produces feedback by relying on 
ad-hoc rules. 


\end{itemize}

\noindent
{\bf Programming by example}
Our work differs from prior programming-by-example (PBE) techniques in two
ways. First, to our knowledge, our work is the first to synthesize
imperative programs with loops. 
Most of the PBE approaches focus on domain-specific languages for
string
transformation~\cite{pbe:string1,pbe:string2,pbe:string3,pbe:string4,pbe:string5},
number transformation~\cite{pbe:number}, 
XML transformation~\cite{pbe:xml},
and extracting relational data~\cite{pbe:flashextract}, etc.  
Several others have studied
synthesis of functional programs
\cite{pbe:escher,pbe:myth,pbe:myth2}.  Second, our algorithm differs
from prior work in that we combine semantic-based static analysis
technology with enumerative program synthesis.  Existing
enumerative synthesis technology used pruning techniques such as type
systems~\cite{pbe:myth,pbe:myth2} and
deductions~\cite{pbe:ds-manipulating}, which are not applicable to our
setting.

\section{Conclusion}

In this paper, we have shown that combining enumerative synthesis and
static analysis is a promising way of synthesizing introductory
imperative programs. The enumerative search allows us to find the
smallest possible, therefore general, program while the
semantics-based static analysis
dramatically accelerates the process in a safe way.
We demonstrated the effectiveness on 30 real programming problems
gathered from online forums.


\bibliographystyle{named}
\bibliography{ref}

\begin{thebibliography}{}

\bibitem[\protect\citeauthoryear{Adam and Laurent}{1980}]{laura:1980}
Anne Adam and Jean-Pierre Laurent.
\newblock Laura, a system to debug student programs.
\newblock {\em Artificial Intelligence}, 15(1-2), November 1980.

\bibitem[\protect\citeauthoryear{Ahmed \bgroup \em et al.\egroup
  }{2013}]{probgen:ijcai13}
Umair~Z. Ahmed, Sumit Gulwani, and Amey Karkare.
\newblock Automatically generating problems and solutions for natural
  deduction.
\newblock In {\em IJCAI}, 2013.

\bibitem[\protect\citeauthoryear{Aho \bgroup \em et al.\egroup
  }{1986}]{dragonbook}
Alfred~V. Aho, Ravi Sethi, and Jeffrey~D. Ullman.
\newblock {\em Compilers: Principles, Techniques, and Tools}.
\newblock Addison-Wesley Longman Publishing Co., Inc., Boston, MA, USA, 1986.

\bibitem[\protect\citeauthoryear{Albarghouthi \bgroup \em et al.\egroup
  }{2013}]{pbe:escher}
Aws Albarghouthi, Sumit Gulwani, and Zachary Kincaid.
\newblock Recursive program synthesis.
\newblock In {\em CAV}, 2013.

\bibitem[\protect\citeauthoryear{Alur \bgroup \em et al.\egroup
  }{2013}]{dfagrading:ijcai13}
Rajeev Alur, Loris D'Antoni, Sumit Gulwani, Dileep Kini, and Mahesh
  Viswanathan.
\newblock Automated grading of dfa constructions.
\newblock In {\em IJCAI}, 2013.

\bibitem[\protect\citeauthoryear{Alvin \bgroup \em et al.\egroup
  }{2014}]{probgen:aaa14}
Chris Alvin, Sumit Gulwani, Rupak Majumdar, and Supratik Mukhopadhyay.
\newblock Synthesis of geometry proof problems.
\newblock In {\em AAAI}, 2014.

\bibitem[\protect\citeauthoryear{Cousot and Cousot}{1977}]{aiframework}
Patrick Cousot and Radhia Cousot.
\newblock Abstract interpretation: A unified lattice model for static analysis
  of programs by construction or approximation of fixpoints.
\newblock In {\em POPL}, 1977.

\bibitem[\protect\citeauthoryear{Farrell \bgroup \em et al.\egroup
  }{1984}]{lisptutor:1984}
Robert~G. Farrell, John~R. Anderson, and Brian~J. Reiser.
\newblock An interactive computer-based tutor for lisp.
\newblock In {\em AAAI}, 1984.

\bibitem[\protect\citeauthoryear{Feser \bgroup \em et al.\egroup
  }{2015}]{pbe:ds-manipulating}
John~K. Feser, Swarat Chaudhuri, and Isil Dillig.
\newblock Synthesizing data structure transformations from input-output
  examples.
\newblock In {\em PLDI}, 2015.

\bibitem[\protect\citeauthoryear{Frankle \bgroup \em et al.\egroup
  }{2016}]{pbe:myth2}
Jonathan Frankle, Peter-Michael Osera, David Walker, and Steve Zdancewic.
\newblock Example-directed synthesis: A type-theoretic interpretation.
\newblock In {\em POPL}, 2016.

\bibitem[\protect\citeauthoryear{Gulwani \bgroup \em et al.\egroup
  }{2011}]{solutiongen:pldi11}
Sumit Gulwani, Vijay~Anand Korthikanti, and Ashish Tiwari.
\newblock Synthesizing geometry constructions.
\newblock In {\em PLDI}, 2011.

\bibitem[\protect\citeauthoryear{Gulwani \bgroup \em et al.\egroup
  }{2014}]{feedback:fse14}
Sumit Gulwani, Ivan Radi\v{c}ek, and Florian Zuleger.
\newblock Feedback generation for performance problems in introductory
  programming assignments.
\newblock In {\em FSE}, 2014.

\bibitem[\protect\citeauthoryear{Gulwani}{2011}]{pbe:string1}
Sumit Gulwani.
\newblock Automating string processing in spreadsheets using input-output
  examples.
\newblock In {\em POPL}, 2011.

\bibitem[\protect\citeauthoryear{Johnson and Soloway}{1984}]{proust:1984}
W.~Lewis Johnson and Elliot Soloway.
\newblock Proust: Knowledge-based program understanding.
\newblock In {\em ICSE}, 1984.

\bibitem[\protect\citeauthoryear{Kaleeswaran \bgroup \em et al.\egroup
  }{2016}]{feedback:fse16}
Shalini Kaleeswaran, Anirudh Santhiar, Aditya Kanade, and Sumit Gulwani.
\newblock Semi-supervised verified feedback generation.
\newblock In {\em FSE}, 2016.

\bibitem[\protect\citeauthoryear{Kim \bgroup \em et al.\egroup }{2016}]{apex}
Dohyeong Kim, Yonghwi Kwon, Peng Liu, I.~Luk Kim, David~Mitchel Perry, Xiangyu
  Zhang, and Gustavo Rodriguez-Rivera.
\newblock Apex: Automatic programming assignment error explanation.
\newblock In {\em OOPSLA}, 2016.

\bibitem[\protect\citeauthoryear{Kini and Gulwani}{2015}]{pbe:string2}
Dileep Kini and Sumit Gulwani.
\newblock Flashnormalize: Programming by examples for text normalization.
\newblock In {\em IJCAI}, 2015.

\bibitem[\protect\citeauthoryear{Le and Gulwani}{2014}]{pbe:flashextract}
Vu~Le and Sumit Gulwani.
\newblock Flashextract: A framework for data extraction by examples.
\newblock In {\em PLDI}, 2014.

\bibitem[\protect\citeauthoryear{Manshadi \bgroup \em et al.\egroup
  }{2013}]{pbe:string4}
Mehdi Manshadi, Daniel Gildea, and James Allen.
\newblock Integrating programming by example and natural language programming.
\newblock In {\em AAAI}, 2013.

\bibitem[\protect\citeauthoryear{Murray}{1989}]{talus:1989}
William~R. Murray.
\newblock {\em Automatic Program DeBugging for Intelligent Tutoring Systems}.
\newblock Morgan Kaufmann Publishers Inc., 1989.

\bibitem[\protect\citeauthoryear{Osera and Zdancewic}{2015}]{pbe:myth}
Peter-Michael Osera and Steve Zdancewic.
\newblock Type-and-example-directed program synthesis.
\newblock In {\em PLDI}, 2015.

\bibitem[\protect\citeauthoryear{Polozov \bgroup \em et al.\egroup
  }{2015}]{probgen:ijcai15}
Oleksandr Polozov, Eleanor O'Rourke, Adam~M. Smith, Luke Zettlemoyer, Sumit
  Gulwani, and Zoran Popovic.
\newblock Personalized mathematical word problem generation.
\newblock In {\em IJCAI}, 2015.

\bibitem[\protect\citeauthoryear{Raza \bgroup \em et al.\egroup
  }{2014}]{pbe:xml}
Mohammad Raza, Sumit Gulwani, and Natasa Milic-Frayling.
\newblock Programming by example using least general generalizations.
\newblock In {\em AAAI}, 2014.

\bibitem[\protect\citeauthoryear{Raza \bgroup \em et al.\egroup
  }{2015}]{pbe:string3}
Mohammad Raza, Sumit Gulwani, and Natasa Milic-Frayling.
\newblock Compositional program synthesis from natural language and examples.
\newblock In {\em IJCAI}, 2015.

\bibitem[\protect\citeauthoryear{Singh and Gulwani}{2012}]{pbe:number}
Rishabh Singh and Sumit Gulwani.
\newblock Synthesizing number transformations from input-output examples.
\newblock In {\em CAV}, 2012.

\bibitem[\protect\citeauthoryear{Singh \bgroup \em et al.\egroup
  }{2012}]{probgen:aaai12}
Rohit Singh, Sumit Gulwani, and Sriram Rajamani.
\newblock Automatically generating algebra problems.
\newblock In {\em AAAI}, 2012.

\bibitem[\protect\citeauthoryear{Singh \bgroup \em et al.\egroup
  }{2013}]{feedback:pldi13}
Rishabh Singh, Sumit Gulwani, and Armando Solar-Lezama.
\newblock Automated feedback generation for introductory programming
  assignments.
\newblock In {\em PLDI}, 2013.

\bibitem[\protect\citeauthoryear{Soloway \bgroup \em et al.\egroup
  }{1981}]{memo2:1981}
Elliot~M. Soloway, Beverly Woolf, Eric Rubin, and Paul Barth.
\newblock Meno-ii: An intelligent tutoring system for novice programmers.
\newblock In {\em IJCAI}. Morgan Kaufmann Publishers Inc., 1981.

\bibitem[\protect\citeauthoryear{Wu and Knoblock}{2015}]{pbe:string5}
Bo~Wu and Craig~A. Knoblock.
\newblock An iterative approach to synthesize data transformation programs.
\newblock In {\em IJCAI}, 2015.

\end{thebibliography}

\end{document}